\documentclass[10pt]{iopart}

\usepackage{graphicx}
\usepackage{cite}

\begin{document}

\title[Subconfiguration-average and level-to-level treatment of electron-impact single ionisation of W$^{14+}$]{Electron-impact single ionisation of W$^\mathbf{14+}$ ions: Subconfiguration-average and level-to-level distorted wave calculations}

\author{F~Jin$^{1,2}$,  A~Borovik, Jr.$^2$, B~Ebinger$^{2,3}$ and S~Schippers$^2$}
\address{$^1$Department of Physics, National University of Defense Technology, Fuyuan Road 1, 410022 Changsha, People's Republic of China}
\address{$^2$I. Physikalisches Institut, Justus-Liebig-Universit\"at Gie\ss{}en, Heinrich-Buff-Ring 16, 35392 Giessen, Germany}
\address{$^3$GSI Helmholtzzentrum f\"{u}r Schwerionenforschung GmbH, Planckstr. 1, 64291 Darmstadt, Germany}
\ead{ftjin@nudt.edu.cn, schippers@jlug.de}
\vspace{10pt}
\begin{indented}
\item[]\today
\end{indented}

\begin{abstract}
The  cross section for electron-impact single ionisation of W$^{14+}$ ions has been calculated by using two different approaches, i.e., the subconfiguration averaged distorted-wave (SCADW) method and the more involved  level-to-level distorted-wave (LLDW) method. Both methods are found to yield very similar results except for the $4d\to 5d$ excitation-autoionisation (EA) channels that straddles the ionisation threshold. Accordingly, a hybrid theoretical cross section where the $4d\to 5d$ EA SCADW cross section is replaced by its LLDW counterpart is in good agreement with the experimental result from an electron-ion crossed-beams experiment. This is in contrast to pure SCADW calculations for W$^{14+}$ and neighbouring charge states which exhibit significant deviations from the experimental near-threshold cross sections of Schury et al.\ [2020 \textit{J.\ Phys.\ B} \textbf{53} 015201].
\end{abstract}

%
%
\submitto{\JPB}
%
%
\ioptwocol

\section{\label{intro}Introduction}
Atomic properties and collision cross sections of tungsten ions are of current interest \cite{Mueller2015b}, since tungsten is used as a plasma-facing material in tokamak facilities \cite{Kallenbach2005}. Inevitably, tungsten ions will thus be abundant in the tokamak plasma and cause strong radiation losses which potentially prevent ignition of the desired thermonuclear reaction. Theoretical plasma modelling is used to predict the abundance and charge balance of tungsten ions in the tokamak plasma. This requires accurate input data from atomic physics, such as cross sections for electron-impact excitation and electron-impact ionisation  or electron-ion recombination. Here, we report on a detailed theoretical study of the cross section for electron-impact single ionisation (EISI) of W$^{14+}$([Kr]$4d^{10}\,4f^{12}\,5s^2$) ions which we compare with recent experimental results \cite{Schury2019}.

Theoretical calculations for EISI of W$^{q+}$ have been carried out also for higher charges states $q$.  Loch \etal~\cite{Loch2005a} used the semirelativistic configuration averaged distorted-wave (CADW) method and published cross sections for $q=9$, 22, 45, 63, 64, and 72. For $q=9$ and at high electron-ion collision energies they found good agreement with the experimental data of Stenke \etal \cite{Stenke1995a}. At energies close to the threshold, however, the theoretical calculations underestimated the measured data significantly. More recently, Pindzola and Loch performed DW and time-dependent close-coupling (TDCC) calculations for EISI of singly charged ions \cite{Pindzola2019a}. Both methods resulted in cross sections that are much larger than the experimental values of Stenke \textit{et al}. Demura \etal \cite{Demura2015} used a statistical approach for the theoretical description of the complex electronic structure of many-electron tungsten ions and presented EISI cross sections for $q= 1-10, 17, 22, 45$, and 63. These calculations reproduce the experimental cross sections for $q=1-10$ \cite{Stenke1995a} and for $q=17$ \cite{Rausch2011} only poorly. A dedicated study on $q=17$ was presented by Zhang and Kwon using a level-to-level distorted wave (LLDW) approach \cite{Zhang2014}. Good agreement between the experimental cross section \cite{Rausch2011} and the theoretical cross section was found when a reasonable assumption for the initial state population in the experiment was made. A similar approach as the one of Loch \etal~\cite{Loch2005a} was used in a joint experimental and theoretical study for $q=19$ by Borovik~Jr.\ \etal\ \cite{Borovik2016} who showed that the inclusion of excitation-autoionisation (EA) involving high-$n$ levels up to $n_\mathrm{max}=23$ leads to better agreement between experiment and theory also close to the threshold as compared to the results of Loch \etal~\cite{Loch2005a} who limited their calculations to $n_\mathrm{max}=8$. The importance of high-$n$ EA channels has also been pointed out in a sequence of theoretical studies for $q=25-27$ by Jonauskas \etal~\cite{Jonauskas2015} and Kynien\.{e} \etal~\cite{Kyniene2015,Kyniene2016}. In addition to sub-configuration averaged distorted wave (SCADW) cross sections for direct ionisation (DI) these authors calculated fine-structure resolved excitation-autoionisation (EA) contributions involving high-$n$ shells using the LLDW method. Unfortunately and in contrast to the present work, there are no experimental data available for benchmarking these calculations.

For the present calculations of EISI of  W$^{14+}$ we employ the LLDW and SCADW methods. It should be noted that the atomic structure of this ion with a [Kr]$4d^{10}\,4f^{12}\,5s^{2}$ ground configuration is of higher complexity than that of the above mentioned W$^{25+}$--W$^{27+}$ ions with only up to three $4f$ electrons outside otherwise closed subshells. For example, EA of W$^{14+}$ associated with $4d\to5d$ ionisation involves a [Kr]$4d^9\,4f^{12}\,5s^2\,5d$ configuration which has two open $d$-shells and one open $f$-shell and which splits into 992 levels. The comparison of our theoretical EISI cross section with the corresponding experimental data \cite{Schury2019} showcases that particularly this significant EA channel which straddles the ionisation threshold can only be accounted for appropriately by a level-to-level approach.

\section{\label{method}Theoretical method}

In general, EISI of an ion can proceed via different routes. The most discussed ionisation processes in the literature (see \cite{Mueller2008a} for a comprehensive overview) are direct ionisation (DI), excitation-autoionisation (EA), and resonant-excitation double-autoionisation (REDA) which can be expressed as
\begin{equation}\label{eq:DI}
e^{-}+A^{q+}\rightarrow A^{(q+1)+}+2e^{-},
\end{equation}
\begin{equation}\label{eq:EA}
e^{-}+A^{q+}\rightarrow [ A^{q+}]^{**}+e^{-}\rightarrow A^{(q+1)+} + 2e^{-},
\end{equation}
and
\begin{eqnarray}\label{eq:REDA}
e^{-}+A^{q+} \rightarrow [ A^{(q-1)+}]^{**} &\rightarrow& [ A^{q+}]^{**}+e^{-}\\ &&\rightarrow A^{(q+1)+} + 2e^{-},
\end{eqnarray}
respectively. The double star superscripts denote autoionising intermediate levels. REDA is a higher-order process which can be important in certain cases \cite{LaGattuta1981,Mueller1988}. Our present calculations show that REDA is not significant for EISI of W$^{14+}$ ions. Therefore, we here do not present any details of our REDA calculations which follow the treatment by Ebinger \etal \cite{Ebinger2019}.

If only DI and EA are accounted  for, the total cross section for EISI from a level $i$ of the ion A$^{q+}$ to a level $j$ of the ion A$^{(q+1)+}$ can be written as
\begin{equation}\label{eq:sigmaij}
\sigma_{ij}(\varepsilon)=\sigma_{ij}^{DI}(\varepsilon)+\sum_k{\sigma_{ik}^{CE}(\varepsilon)B_{kj}},
\end{equation}
where $\sigma_{ij}^{DI}(\varepsilon)$ is the DI cross section at the incident electron energy $\varepsilon$, and $\sigma_{ik}^{CE}(\varepsilon)$ is the  cross section for electron-impact excitation from level $i$ to level $k$ of the ion A$^{q+}$. In Eq.~\ref{eq:sigmaij}, $B_{kj}$ is the branching ratio for autoionisation which is calculated as
\begin{equation}\label{eq:br}
    B_{kj}=\frac{A_{kj}^{a}+\sum_n{A^r_{kn}B_{nj}}}{\sum_m{A_{km}^{a}+\sum_n{A_{kn}^{r}}}}.
\end{equation}
Here, $A_{kj}^a$ is the rate for the Auger transition from the level $k$ of the ion $A^{q+}$ to the level $j$ of the ion $A^{(q+1)+}$. Likewise, $A^r_{kn}$ denotes the rate for a radiative transition from the level $k$ to a level $n$ of the ion A$^{q+}$. The second term in the numerator accounts for transition cascades involving energetically lower autoionising levels. In general, cascades do not have a large impact on the branching ratios. Nevertheless, the present calculations account for all energetically possible cascade steps.  All required atomic quantities were calculated with the Dirac-Fock-Slater method as implemented in the Flexible Atomic Code (FAC) \cite{gu2008} which also provides the fully relativistic SCADW method for the calculation of DI cross sections.

Here, we consider DI of a $5s$, $4f$, and $4d$ electron from the ground configuration\footnote{For brevity, the electrons in the filled K, L and M inner shells are not included in the notation.} of W$^{14+}$, i.e.,
\begin{eqnarray}
\lefteqn{e+4s^{2}\,4p^{6}\,4d^{10}\,4f^{12}\,5s^{2}\rightarrow}\nonumber\\
&&\hspace*{1cm} 2e+
\left\lbrace
\begin{array}{l}
4s^{2}\,4p^{6}\,4d^{10}\,4f^{12}\,5s\\
4s^{2}\,4p^{6}\,4d^{10}\,4f^{11}\,5s^{2}\\
4s^{2}\,4p^{6}\,4d^{9\phantom{0}}\,4f^{12}\,5s^{2}
\end{array}
\label{cichannel}
\right\rbrace
.
\end{eqnarray}
The thresholds for DI of a $4p$ and a $4s$ electron occur at 726.88 eV and at 901.60 eV, respectively, i.e., beyond the threshold for double ionisation at 686.9 eV \cite{ASD2019}. In our calculation, DI of a $4p$ or a $4s$ electron is small and found to contribute primarily to double ionisation and, thus, is disregarded in the present EISI cross section.

EA processes are initiated by the excitation of an electron from the $4d$, $4p$, or $4s$ subshell to a subshell $nl$:
\begin{eqnarray}
\lefteqn{e+4s^{2}\,4p^{6}\,4d^{10}\,4f^{12}\,5s^{2}\rightarrow}\nonumber\\
&&\hspace*{1cm} e+\left\lbrace
\begin{array}{l}
4s^{2}\,4p^{6}\,4d^{9\phantom{0}}\,4f^{12}\,5s^{2}\,nl\\
4s^{2}\,4p^{5}\,4d^{10}\,4f^{12}\,5s^{2}\,nl\\
4s\phantom{^{2}}\,4p^{6}\,4d^{10\,}4f^{12}\,5s^{2}\,5l
\end{array}
\right\rbrace
\label{eachannel}
\end{eqnarray}
For $4d$ and $4p$ excitations we consider subshells with $n\leq25$  and $n\leq9$, respectively, and with $l\leq8$ in both cases. Our calculations suggest that the total EISI cross section has converged when these maximum $n$ and $l$ values are used. We exclude excitations to the $4f$ subshell since the resulting levels do not autoionise.  $5s\to nl$ excitations lead to autoionising levels for $n\geq24$. However, the resulting contributions to the total EISI cross section are negligible and are, therefore, not included in the present calculations, either.

Previous large-scale EA calculations, e.g., for Cu-like ions have revealed that the configuration interaction (CI) has a significant influence on the computed EA cross sections \cite{Mitnik1996a}. Our SCADW calculations account for CI between all excited configurations with the size of the CI matrices being about $50,000\times50,000$. For the LLDW calculations a full treatment of CI would require much larger matrices and, consequently, has not been possible because of computer-memory constraints. Nevertheless, we were able to account for CI within the same Rydberg series. For the particularly important $4d\rightarrow 5d$ EA channel we have included CI between all $4s^2\,4p^6\,4d^9\,4f^{12}\,5s^2\,nd$ configurations with $5\leq n\leq n_\mathrm{max}$. In order to check for convergence we have performed calculations for $n_\mathrm{max}=12$, 18, and 25 the results of which are virtually identical.

\section{\label{result}Results and discussions}

\begin{table}
	\caption{\label{TABLE1}Degeneracies $g$ and energies of the subconfigurations belonging to the W$^{14+}$, W$^{15+}$, and W$^{16+}$ ground configurations.}
\begin{indented}
	\item[]	
		\begin{tabular}{lclrr}\br
			Ion        & No.\ & Subconfiguration                         & $g$  & Energy (eV) \\
			\mr
			W$^{14+}$  & 0 & $4f_{5/2}^{6}\,4f_{7/2}^{6}\,5s_{1/2}^2$     & 28   &   0        \\
			& 1 & $4f_{5/2}^{5}\,4f_{7/2}^{7}\,5s_{1/2}^2$     & 48   &   2.84     \\
			& 2 & $4f_{5/2}^{4}\,4f_{7/2}^{8}\,5s_{1/2}^2$     & 15   &   4.45     \\
			W$^{15+}$  & 3 & $4f_{5/2}^{6}\,4f_{7/2}^{5}\,5s_{1/2}^2$     & 56   & 323.40     \\
			& 4 & $4f_{5/2}^{5}\,4f_{7/2}^{6}\,5s_{1/2}^2$     & 168  & 326.85     \\
			& 5 & $4f_{5/2}^{4}\,4f_{7/2}^{7}\,5s_{1/2}^2$     & 120  & 329.01     \\
			& 6 & $4f_{5/2}^{3}\,4f_{7/2}^{8}\,5s_{1/2}^2$     & 20   & 329.90     \\
			W$^{16+}$  & 7 & $4f_{5/2}^{6}\,4f_{7/2}^{5}\,5s_{1/2}$       & 112  & 684.85     \\
			& 8 & $4f_{5/2}^{5}\,4f_{7/2}^{6}\,5s_{1/2}$       & 336  & 688.36     \\
			& 9 & $4f_{5/2}^{4}\,4f_{7/2}^{7}\,5s_{1/2}$       & 240  & 690.59     \\
			&10 & $4f_{5/2}^{3}\,4f_{7/2}^{8}\,5s_{1/2}$       & 40   & 691.55  \\
\br
		\end{tabular}
	\end{indented}
\end{table}

The ground configuration of W$^{14+}$ is [Kr]$4d^{10}4f^{12}5s^2$, which splits into three subconfigurations in the fully relativistic SCADW calculation. The subconfiguration energies of W$^{14+}$ as well as W$^{15+}$ and W$^{16+}$ are shown in table \ref{TABLE1}. The calculated energies of the first two excited subconfigurations of W$^{14+}$ are only 2.84 eV and 4.45 eV above the ground subconfiguration. The calculated single and double ionisation potentials are 323.40 eV and 684.85 eV, respectively. These values agree reasonably well with the corresponding NIST values of 325.3 eV and 686.9 eV \cite{ASD2019}.

\begin{figure}
\includegraphics[width=\columnwidth]{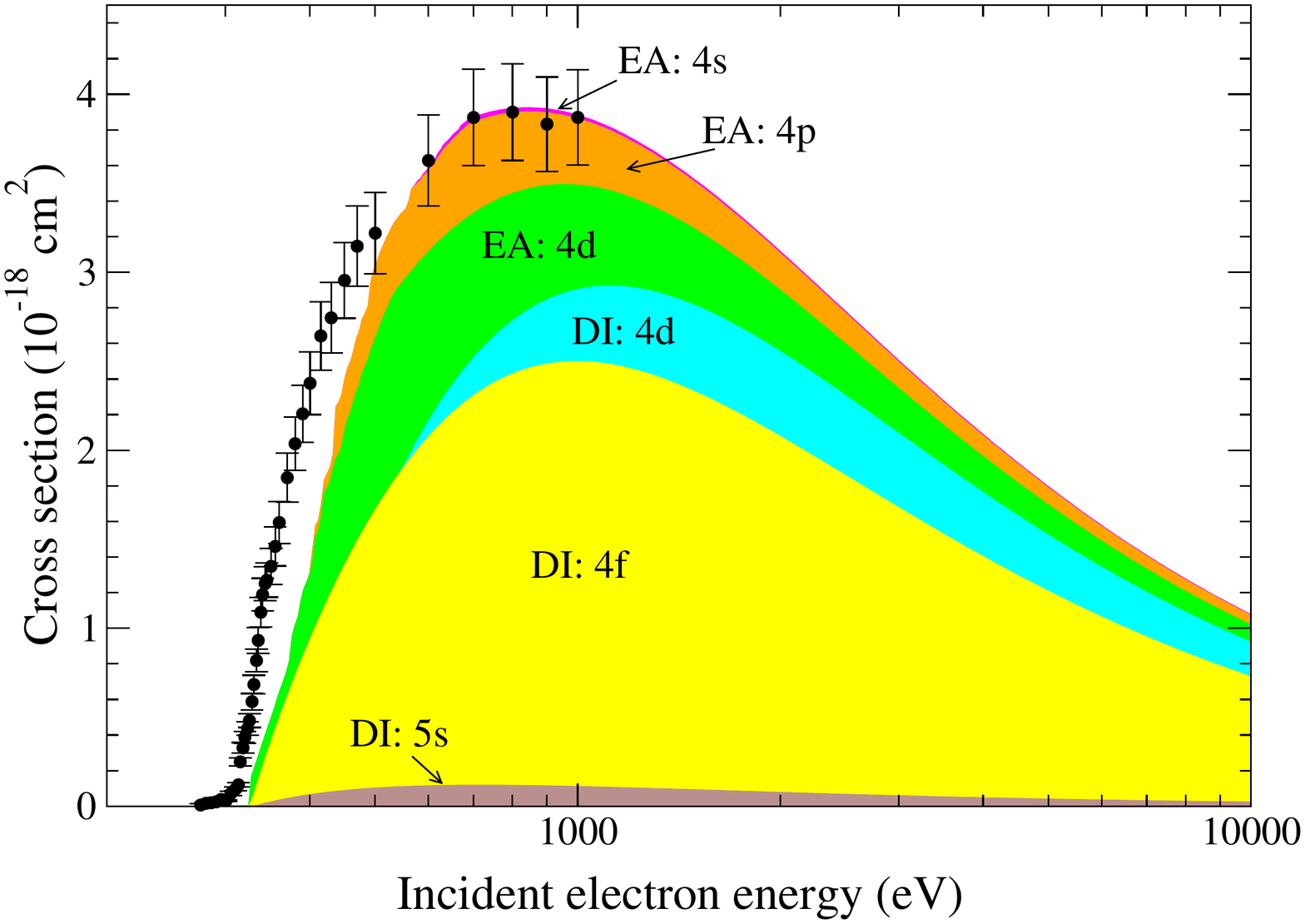}
\caption{\label{fig1} Electron-impact single-ionisation (EISI) cross sections of W$^{14+}$. The black filled circles with error bars are experimental data measured by D. Schury \etal \cite{Schury2019}. The shaded curves result from the present SCADW calculations for EISI of the W$^{14+}$ ground configuration. The calculations comprise DI of  $5s$, $4f$, and $4d$ electrons as well as EA involving excitations of the $4d$, $4p$ and $4s$ subshells.}
\end{figure}

Figure \ref{fig1} shows the EISI cross section of the ground configuration of W$^{14+}$ as resulting from a pure SCADW calculation in comparison with the experimental cross section of Schury \etal~\cite{Schury2019}. Within this approach, which was also already employed by these authors,  the cross section for a configuration is obtained as the statistical average of the cross sections for the pertinent subconfigurations. In figure~\ref{fig1}, the DI and EA cross sections of each channel are plotted separately. The major contribution to the DI cross section comes from the direct knock-out of a $4f$ electron. The cross sections for DI of a $4d$ or a $5s$ electron are much smaller, but still significant.  The major contributions to the EA cross section are associated with $4d$ and $4p$ excitations. The contributions by $4s$ excitations are negligibly small.

\begin{figure}
\includegraphics[width=\columnwidth]{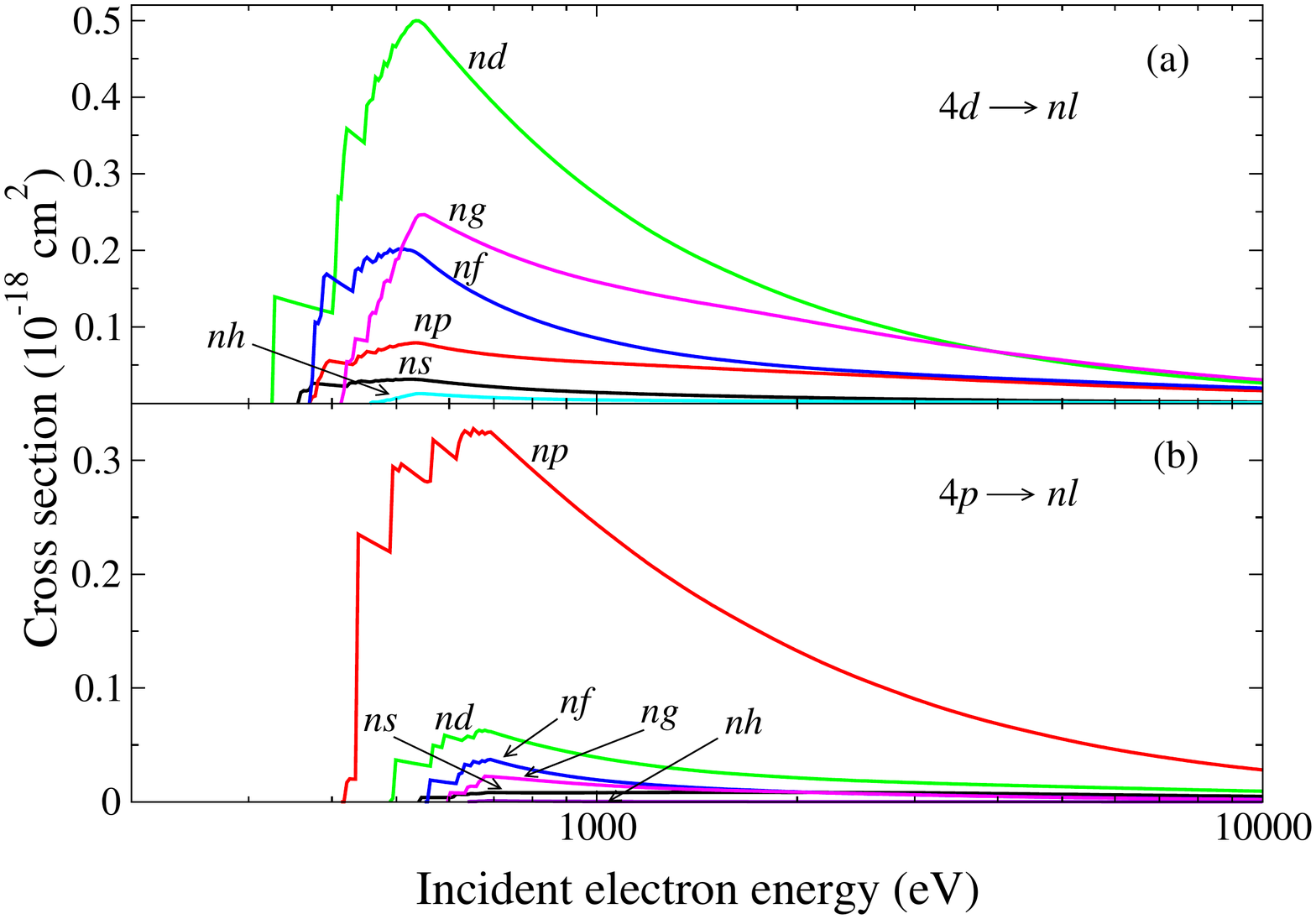}
\caption{\label{fig2} EA contributions to the SCADW cross section for EISI of W$^{14+}$ ions. (a) EA cross sections associated with $4d\rightarrow nl$ excitations with $5\leq n \leq 25$. Individual curves are plotted for $0\leq l\leq 5$. (b) EA cross sections associated with $4p\rightarrow nl$ excitations with $5\leq n \leq 9$ and $0\leq l\leq 5$.}
\end{figure}

The dominant EA contributions are displayed in figure~\ref{fig2} in more detail. The most important EA channels are those associated with $4d\rightarrow nd$ and $4p\rightarrow np$ excitations. The EA cross sections tend to become smaller with increasing angular momentum $l$ of the excited $nl$ electron. We find that the EA cross section has practically converged if all  $4d\rightarrow nl$ and $4p\rightarrow n'l$ excitations with  $5\leq n \leq 25$, $5\leq n' \leq 9$, and $0\leq l \leq 8$ are included in the calculation.

Configurations with an open $4f$ subshell can split into large numbers of levels. In the present SCADW calculations, ten thousands of subconfigurations are included which potentially split into millions of levels. Therefore, a full level-to-level calculation which additionally accounts for all configuration interactions is practically not feasible. Nevertheless, we have performed some investigations into the differences between the configuration averaged  and the level-to-level approaches. Figure \ref{fig1} reveals that the SCADW cross section is significantly lower than the experimental cross section for energies from above the ionisation threshold at $\sim$325 eV up to about 500 eV. In the following we argue that this discrepancy is largely caused by an inaccurate SCADW treatment of the  $4d\rightarrow 5d$ EA channel, which is the EA channel closest to the ionisation threshold.

\begin{table}
	\caption{\label{TABLE2} Degeneracies $g$ and energies of the subconfigurations of the $4d^{9}\,4f^{12}\,5s^{2}\,5d$ configuration. The numeration of the subconfigurations is continued from table~\ref{TABLE1}.}
\begin{indented}
	\item[]	
		\begin{tabular}{clcr}\br
			No.\ & Subconfiguration                                                          & $g$  & Energy (eV) \\
			\mr
			11     & $4d_{3/2}^{4}\,4d_{5/2}^{5}\,4f_{5/2}^{6}\,4f_{7/2}^{6}\,5s_{1/2}^{2}\,5d_{3/2}$   & 672  & 308.31  \\
			12     & $4d_{3/2}^{4}\,4d_{5/2}^{5}\,4f_{5/2}^{6}\,4f_{7/2}^{6}\,5s_{1/2}^{2}\,5d_{5/2}$   & 1008 & 311.64  \\
			13     & $4d_{3/2}^{4}\,4d_{5/2}^{5}\,4f_{5/2}^{5}\,4f_{7/2}^{7}\,5s_{1/2}^{2}\,5d_{3/2}$   & 1152 & 313.13  \\
			14     & $4d_{3/2}^{4}\,4d_{5/2}^{5}\,4f_{5/2}^{5}\,4f_{7/2}^{7}\,5s_{1/2}^{2}\,5d_{5/2}$   & 1728 & 316.08  \\
			15     & $4d_{3/2}^{4}\,4d_{5/2}^{5}\,4f_{5/2}^{4}\,4f_{7/2}^{8}\,5s_{1/2}^{2}\,5d_{3/2}$   & 360  & 316.69  \\
			16     & $4d_{3/2}^{4}\,4d_{5/2}^{5}\,4f_{5/2}^{4}\,4f_{7/2}^{8}\,5s_{1/2}^{2}\,5d_{5/2}$   & 540  & 319.26  \\
			17     & $4d_{3/2}^{3}\,4d_{5/2}^{6}\,4f_{5/2}^{6}\,4f_{7/2}^{6}\,5s_{1/2}^{2}\,5d_{3/2}$   & 448  & 325.34  \\
			18     & $4d_{3/2}^{3}\,4d_{5/2}^{6}\,4f_{5/2}^{4}\,4f_{7/2}^{8}\,5s_{1/2}^{2}\,5d_{3/2}$   & 240  & 325.41  \\
			19     & $4d_{3/2}^{3}\,4d_{5/2}^{6}\,4f_{5/2}^{5}\,4f_{7/2}^{7}\,5s_{1/2}^{2}\,5d_{3/2}$   & 768  & 326.01  \\
			20     & $4d_{3/2}^{3}\,4d_{5/2}^{6}\,4f_{5/2}^{4}\,4f_{7/2}^{8}\,5s_{1/2}^{2}\,5d_{5/2}$   & 360  & 327.23  \\
			21     & $4d_{3/2}^{3}\,4d_{5/2}^{6}\,4f_{5/2}^{6}\,4f_{7/2}^{6}\,5s_{1/2}^{2}\,5d_{5/2}$   & 672  & 327.92  \\
			22     & $4d_{3/2}^{3}\,4d_{5/2}^{6}\,4f_{5/2}^{5}\,4f_{7/2}^{7}\,5s_{1/2}^{2}\,5d_{5/2}$   & 1152 & 328.21  \\
\br
		\end{tabular}
	\end{indented}
\end{table}

Table \ref{TABLE2} lists the subconfiguration energies of the $4d^{9}\,4f^{12}\,5s^{2}\,5d$ configuration, which is the final configuration of the $4d\rightarrow 5d$ EA channel. Apparently,  the subconfiguration energies which vary from 308.31 eV to 328.21 eV straddle the ionisation threshold with half of the subconfigurations being below and half above. In addition, also the number of W$^{15+}$ subconfigurations which energetically can be reached from the $4d^{9}\,4f^{12}\,5s^{2}\,5d$ configuration by an Auger transition is limited.  Referring to the numeration of the subconfigurations in Tables \ref{TABLE1} and \ref{TABLE2}, the only possible Auger transitions are $17\rightarrow 3$, $21\rightarrow 4$,  $21\rightarrow 3$ and $22\rightarrow 4$. The remaining energetically allowed transitions, i.e., the transitions $18\rightarrow 3$, $19\rightarrow 3$, $20\rightarrow 3$, $20\rightarrow 4$, and $22\to3$ are excluded in the SCADW model because these would involve a spin flip of a $4f$ electron.

These limitations and the finding that the  $4d^{9}\,4f^{12}\,5s^{2}\,5d$ configuration straddles the ionisation threshold call for a more detailed level-to-level treatment of the $4d\rightarrow 5d$ EA channel. The 13 levels of the ground configuration of W$^{14+}$ and their energies as resulting from our LLDW calculation are listed in table \ref{TABLE3}. The energies extend up to 16.56 eV. This energy is much larger than the largest subconfiguration energy  of 4.45~eV (table~\ref{TABLE1}). The ground configuration of W$^{15+}$ splits into 41 levels ranging from 321.98 to 344.53~eV. Due to the complexity of the $4f$ coupling, the $4d^{9}\,4f^{12}\,5s^{2}\,5d$ configuration splits into 992 levels. Their energies vary across a much larger range of 294.55--353.15~eV as compared to the subconfiguration range of 308.31--328.21~eV. 420 of these levels are beyond the ionisation threshold.

\begin{table*}
	\caption{\label{TABLE3} Calculated energies of the levels belonging to the W$^{14+}$ and W$^{15+}$ ground configurations.}
	\begin{indented}
	\item[]	\begin{tabular}{cclcr}\br
			Ion & Index & Level                                                         &$J$ & Energy (eV) \\
			\mr
			W$^{14+}$ & 0     & [$4f_{5/2}^{6}(0)\,4f_{7/2}^{6}(6)]_{6}\,5s_{1/2}^{2}$      & 6  & 0      \\
			          & 1     & [$4f_{5/2}^{6}(0)\,4f_{7/2}^{6}(4)]_{4}\,5s_{1/2}^{2}$      & 4  & 1.16   \\
			          & 2     & [$4f_{5/2}^{5}(5/2)\,4f_{7/2}^{7}(7/2)]_{5}\,5s_{1/2}^{2}$  & 5  & 2.09   \\
			          & 3     & [$4f_{5/2}^{5}(5/2)\,4f_{7/2}^{7}(7/2)]_{4}\,5s_{1/2}^{2}$  & 4  & 3.04   \\
			          & 4     & [$4f_{5/2}^{5}(5/2)\,4f_{7/2}^{7}(7/2)]_{3}\,5s_{1/2}^{2}$  & 3  & 3.47   \\
			          & 5     & [$4f_{5/2}^{6}(0)\,4f_{7/2}^{6}(2)]_{2}\,5s_{1/2}^{2}$      & 2  & 3.48   \\
			          & 6     & [$4f_{5/2}^{4}(4)\,4f_{7/2}^{8}(0)]_{4}\,5s_{1/2}^{2}$      & 4  & 5.03   \\
			          & 7     & [$4f_{5/2}^{6}(0)\,4f_{7/2}^{6}(2)]_{2}\,5s_{1/2}^{2}$      & 2  & 6.42   \\
			          & 8     & [$4f_{5/2}^{5}(5/2)\,4f_{7/2}^{7}(7/2)]_{6}\,5s_{1/2}^{2}$  & 6  & 7.58   \\
			          & 9     & [$4f_{5/2}^{6}(0)\,4f_{7/2}^{6}(0)]_{0}\,5s_{1/2}^{2}$      & 0  & 8.10   \\
			          & 10    & [$4f_{5/2}^{5}(5/2)\,4f_{7/2}^{7}(7/2)]_{1}\,5s_{1/2}^{2}$  & 1  & 8.46   \\
			          & 11    & [$4f_{5/2}^{4}(2)\,4f_{7/2}^{8}(0)]_{2}\,5s_{1/2}^{2}$      & 2  & 8.90   \\
			          & 12    & [$4f_{5/2}^{4}(0)\,4f_{7/2}^{8}(0)]_{0}\,5s_{1/2}^{2}$      & 0  & 16.56  \\
			W$^{15+}$ & 13    & [$4f_{5/2}^{6}(0)\,4f_{7/2}^{5}(15/2)]_{15/2}\,5s_{1/2}^{2}$  & 15/2  & 321.98   \\
			          & 14    & [$4f_{5/2}^{5}(5/2)\,4f_{7/2}^{6}(6)]_{13/2}\,5s_{1/2}^{2}$   & 13/2  & 323.95   \\
			          & 15    & [$4f_{5/2}^{6}(0)\,4f_{7/2}^{5}(11/2)]_{11/2}\,5s_{1/2}^{2}$  & 11/2  & 324.62   \\
			          & ...   & ...                                                       & ...   & ...      \\
			          & 55    & [$4f_{5/2}^{4}(0)\,4f_{7/2}^{7}(7/2)]_{7/2}\,5s_{1/2}^{2}$    & 7/2   & 344.53   \\
\br
		\end{tabular}
	\end{indented}
\end{table*}

\begin{figure}
\includegraphics[width=\columnwidth]{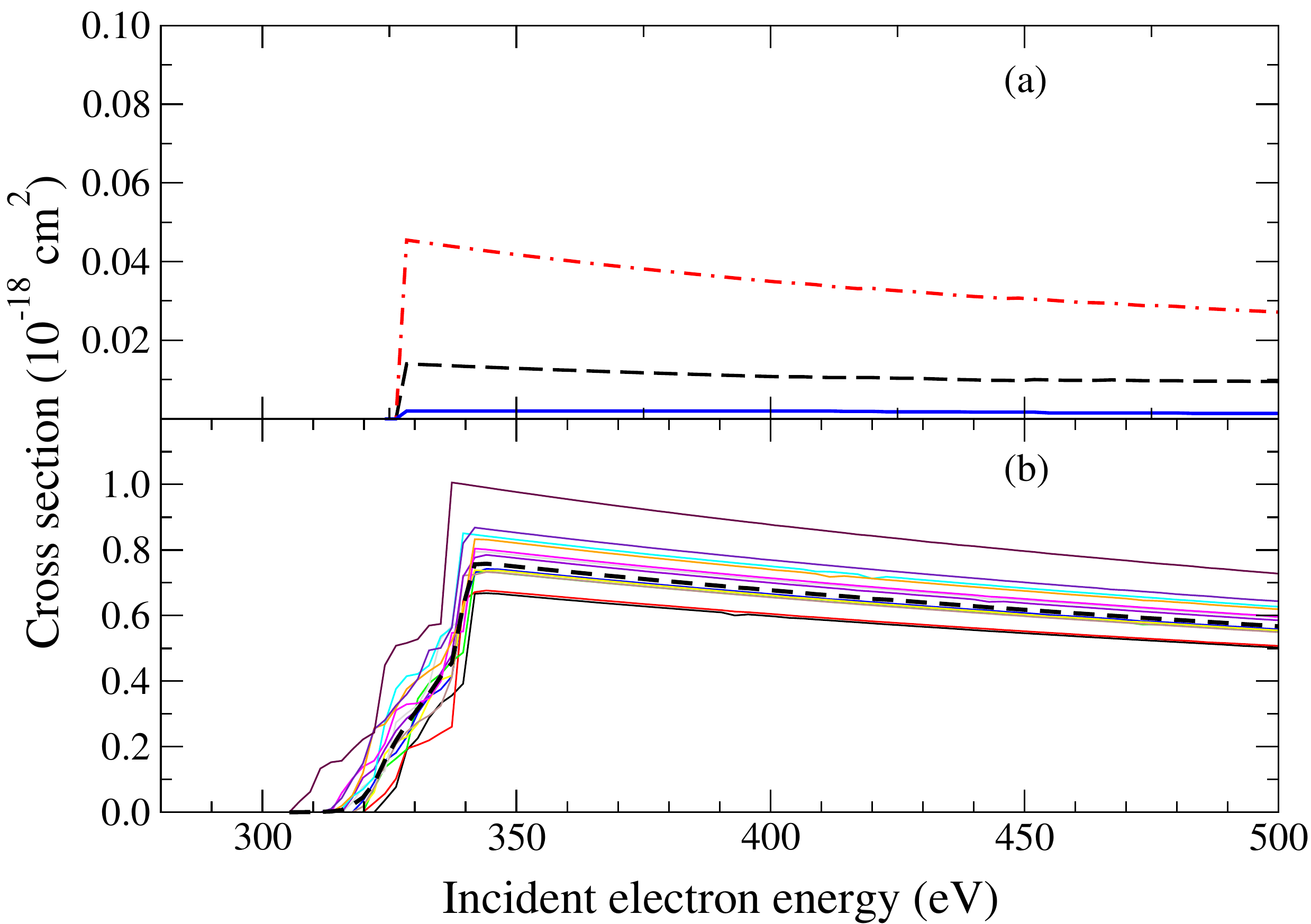}
\caption{\label{fig3} The $4d$$\rightarrow$$5d$ EA cross sections of the ground configuration $4d^{10}\,4f^{12}\,5s^2$. (a) SCADW calculation: The dash-dotted (red) line is the EA cross section of subconfiguration 0 in in table~\ref{TABLE1} and the solid (blue) line is the EA cross section of subconfiguration 1. The dashed (black) line represents the weighted average of these cross sections. (b) LLDW calculation: The coloured solid lines are the EA cross sections of the individual levels from table~\ref{TABLE2}. The thick dashed (black) line represents the weighted average of these. In order to minimise the effort for the comparison between SCADW and LLDW calculations all branching ratios from equation \ref{eq:EA} were set to unity for the generation of this figure.}
\end{figure}

\begin{figure}
\includegraphics[width=\columnwidth]{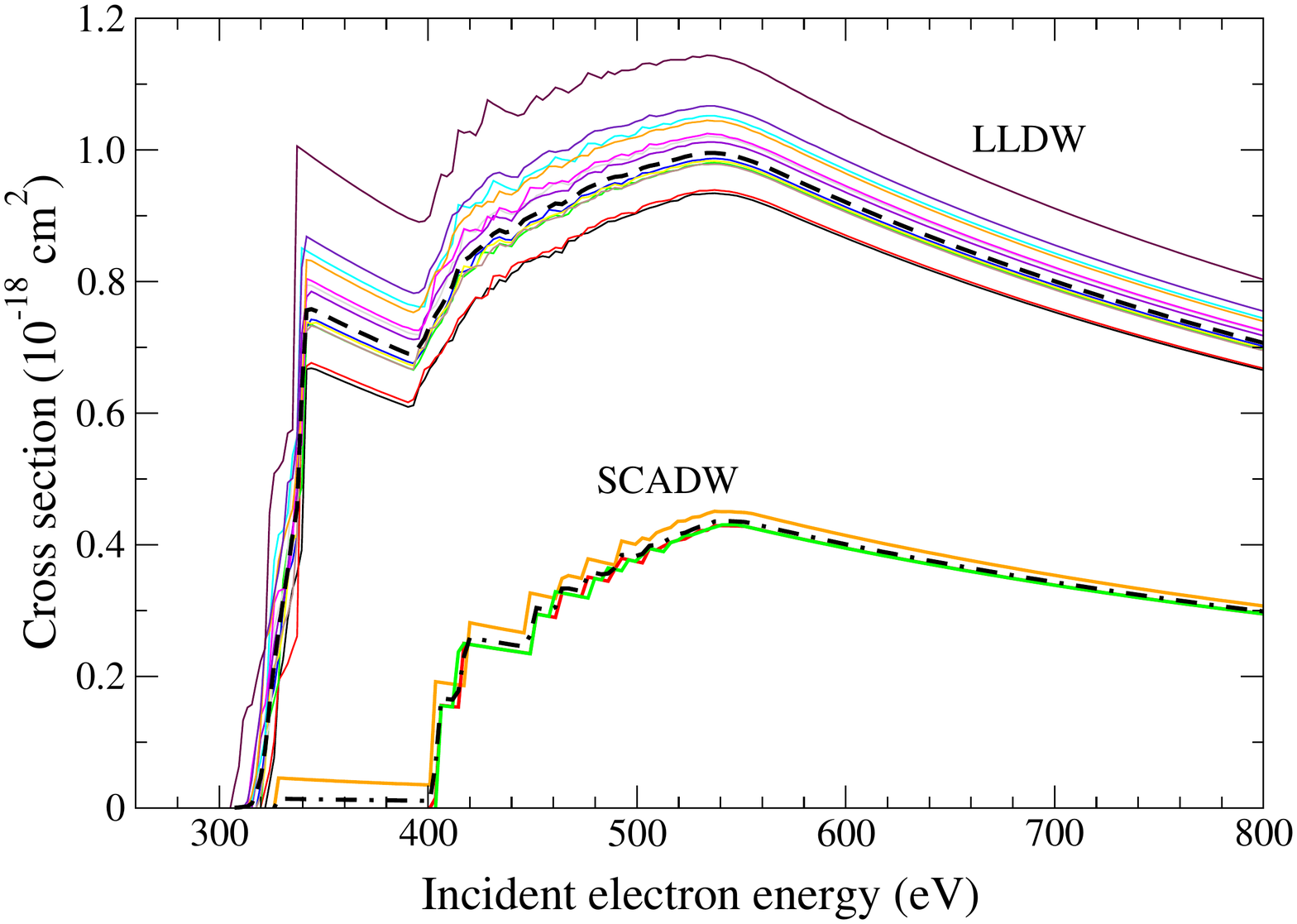}
\caption{\label{fig4} $4d\rightarrow nd$  EA cross sections comprising contributions from $5\leq n\leq25$ for all W$^{14+}$ initial SCADW subconfigurations and initial LLDW levels from tables \ref{TABLE1} and \ref{TABLE3}, respectively. The black dash-dotted and dashed lines represent the weighted averages of SCADW and LLDW cross sections, respectively. As in figure \ref{fig3} all branching ratios  from equation \ref{eq:EA} were set to unity in order to minimise the effort for the comparison.}
\end{figure}

\begin{figure}
\includegraphics[width=\columnwidth]{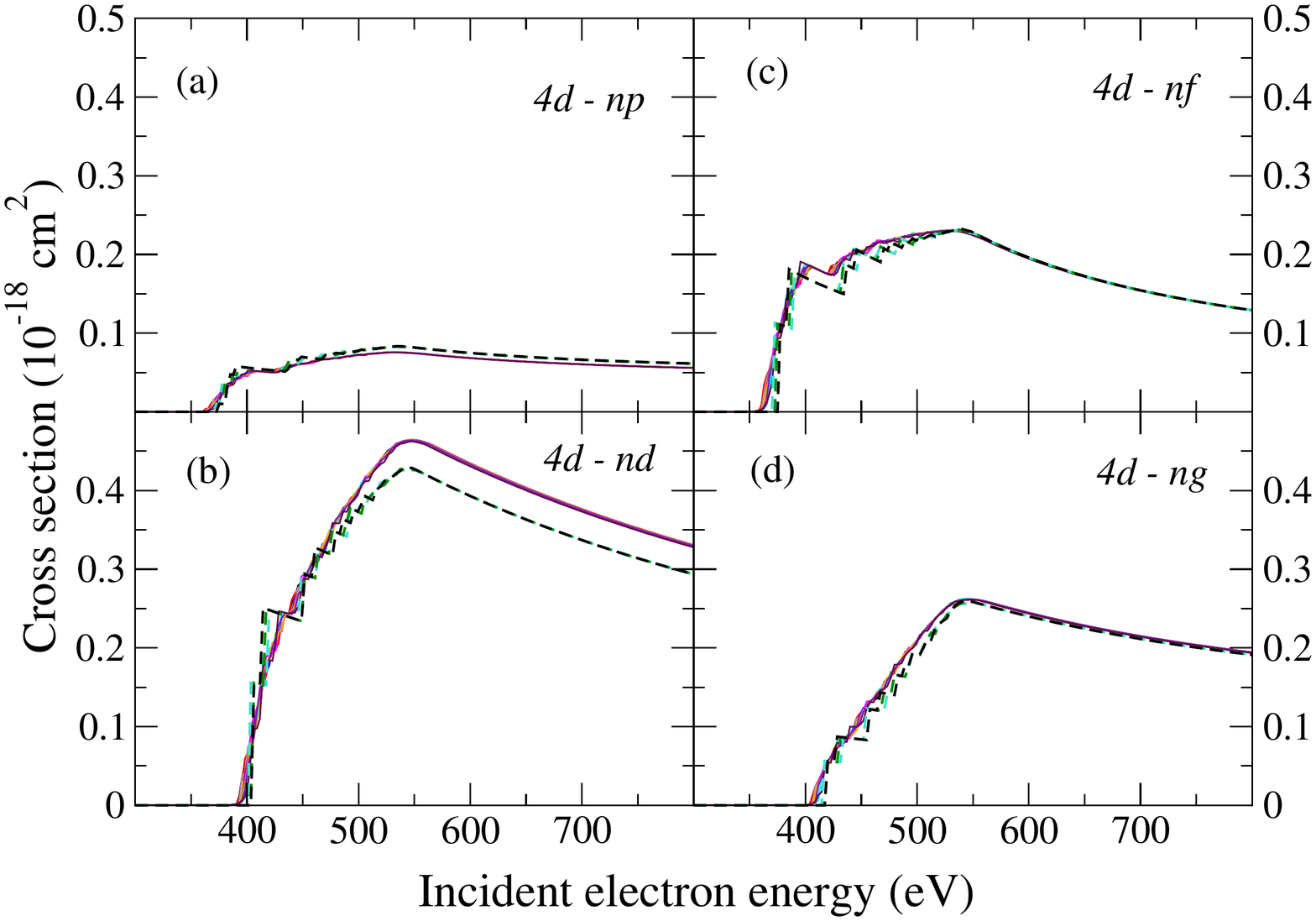}
\caption{\label{fig5} $4d\rightarrow nl$ EA cross sections for different values of the angular momentum quantum number $l$ and for $n_\mathrm{min}\leq n\leq25$ with $n_\mathrm{min}=6(5)$ in case of panels a and b (c and d). The dashed and solids lines are the SCADW and LLDW results, respectively. As in figure \ref{fig3} all branching ratios  from equation \ref{eq:EA} were set to unity in order to minimise the effort for the comparison.}
\end{figure}

\begin{figure}
\includegraphics[width=\columnwidth]{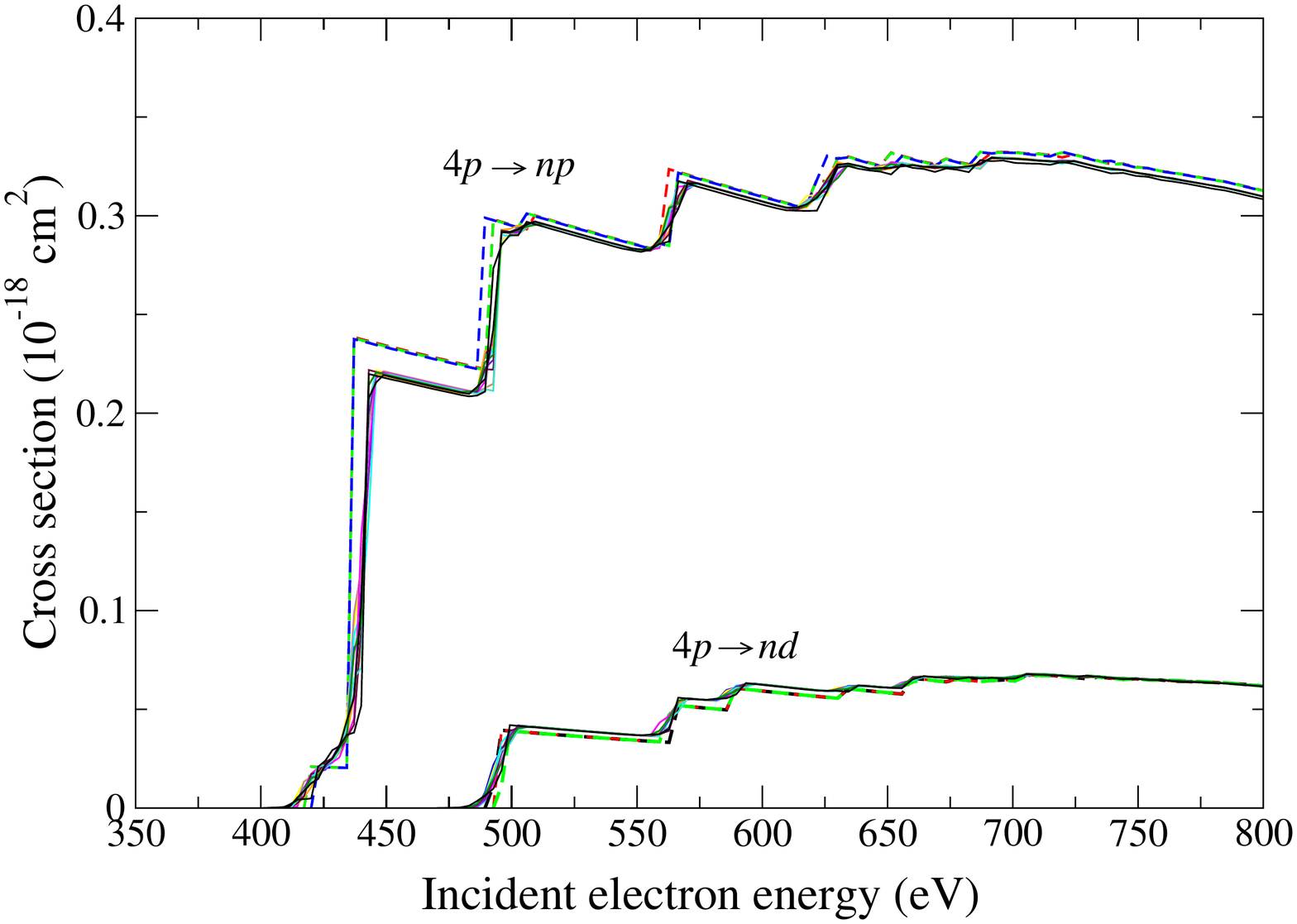}
\caption{\label{fig6} $4p\rightarrow np$ and $4p\rightarrow nd$ EA cross sections comprising contributions from  $5\leq n\leq 9$. In each group dashed and solid lines represent SCADW and LLDW results, respectively. As in figure \ref{fig3} all branching ratios  from equation \ref{eq:EA} were set to unity in order to minimise the effort for the comparison.}
\end{figure}

Figure \ref{fig3} shows the comparison between the SCADW and LLDW $4d\rightarrow 5d$ EA cross sections. The dashed-dotted (red) and the solid (blue) line in figure \ref{fig3}a are the EA cross sections for the subconfigurations 0 and 1 (cf.~table~\ref{TABLE1}) and the dashed (black) line is the weighted average of these two cross sections. Subconfiguration 2 does not contribute to EA since it does not autoionise within the SCADW model. In order to minimise the effort for the comparison between SCADW and LLDW calculations in figure \ref{fig3} and in the subsequent figures \ref{fig4}--\ref{fig6} all branching ratios from equation \ref{eq:EA} were set to unity. This is expected not to introduce significant errors since the branching ratios that have been calculated with the SCADW method are all found to be very close to one. We like to mention already here that the branching ratios were explicitly calculated for all contributions to our final result presented below.

The LLDW cross sections shown in figure~\ref{fig3}b are dramatically different. The ionisation onset is at a considerably lower energy as compared to the SCADW calculation. This is primarily due to EA of level 12 in table~\ref{TABLE3} which has a much lower ionisation energy than any of the subconfigurations in table~\ref{TABLE1}. Most notably, the averaged EA cross section is more than a factor of $\sim$50 larger in the LLDW calculation than in the SCADW model. The difference between the SCADW and LLDW $4d\to nd$ EA results becomes smaller at higher energies where excitations to higher $nd$ subshells contribute in addition. Nevertheless, it still amounts to more than a factor of 2 (figure \ref{fig4}).

Such a large difference between SCADW and LLDW cross sections does not occur in any other EA channel as is shown in  figure~\ref{fig5}. This figure presents angular-momentum-resolved comparisons of the SCADW and LLDW  $4d\rightarrow nl$ EA cross sections for $1\leq l \leq 4$ and $n_\mathrm{min} \leq n \leq 25$. For $4d\rightarrow np$ EA the minimal principal quantum number is  $n_\mathrm{min}=6$ since $n=5$ is energetically not possible. For $4d\rightarrow nd$ EA the same value was used since the $4d\rightarrow 5d$ contribution from figure~\ref{fig3} is deliberately excluded from this comparison. For the higher $l$ contributions $n_\mathrm{min}=5$ has been used. The $4d\rightarrow ns$ and $4d\rightarrow nh$ EA channels are not plotted because these contribute only marginally to the overall EA cross section. We note, that the comparisons presented in figure~\ref{fig5} reveal only minor differences between the SCADW and LLDW curves when the $4d\rightarrow 5d$ EA channel is disregarded.

Figure \ref{fig6} presents the comparison between SCADW and LLDW calculations for EA via $4p\rightarrow np$ and $4p\rightarrow nd$ excitations. As compared to $4d$ EA the EA cross sections for the $4p$ subshell are smaller. However, the EA steps are larger due to the larger differences in excitation energies. In general, the SCADW and LLDW results agree well for the $4p$ EA channels, except for a minor difference in the $4p\rightarrow np$ channels below 500 eV.

\begin{figure}
\includegraphics[width=\columnwidth]{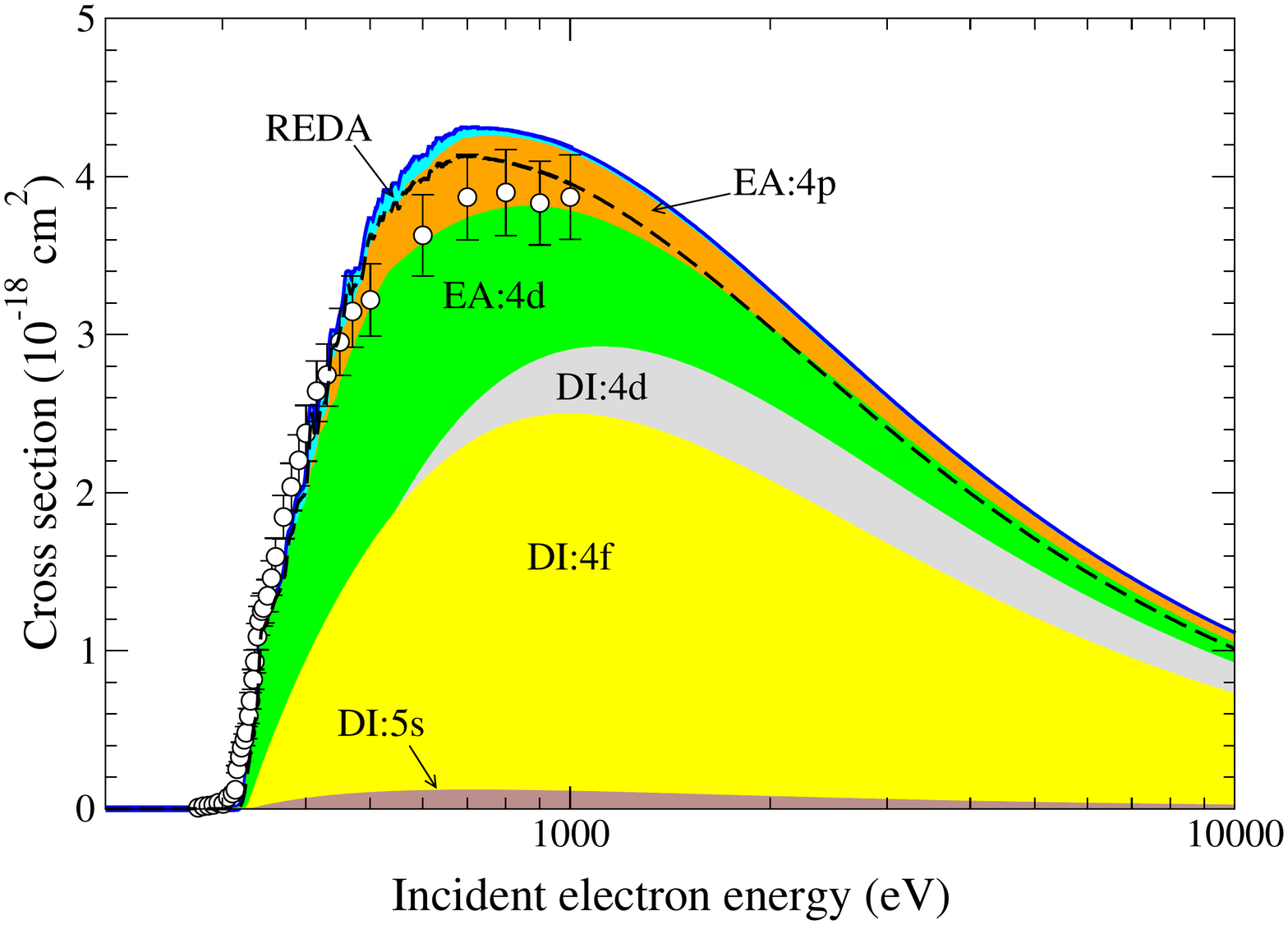}
\caption{\label{fig7} Comparison between the hybrid SCADW/LLDW (see text) total cross section (full blue line) for electron-impact single ionisation of W$^{14+}$ and the experimental cross sections (symbols) of Schury \etal \cite{Schury2019}. The shaded areas illustrate the individual contributions of the various ionisation processes to the total theoretical cross section. The $4s$ EA contribution is too small to be visible on the scale of the figure. The dashed line represents the total cross section from a calculation where the orbitals were optimised on the ground-configuration of the ionised W$^{15+}$ ion.}
\end{figure}

As discussed up to here, the SCADW and LLDW calculations exhibit large differences only for the $4d\rightarrow 5d$ EA cross sections while there is reasonable agreement for the other EA channels where the excited configurations do not straddle the ionisation threshold. Furthermore, there is also good agreement between the SCADW and LLDW cross sections for direct ionisation. Therefore, as our final total cross section for EISI of W$^{14+}$ ions we use a hybrid result which is mostly based on our SCADW calculations including REDA but with the SCADW $4d\rightarrow 5d$ EA cross section replaced by the corresponding LLDW result with the explicit calculation of all branching ratios from equation \ref{eq:EA}. This hybrid cross section is displayed in figure \ref{fig7}. As compared with the pure SCADW calculation presented in figure \ref{fig1}, the agreement between the theoretical and experimental data is greatly improved, in particular, in the energy range from the ionisation threshold up to 500 eV.

Figure \ref{fig7} reveals that REDA processes do not play a significant role in EISI of W$^{14+}$ ions. The REDA cross section is smaller than the experimental error bars. In the vicinity of the cross-section maximum, the total theoretical cross section is significantly larger than the experimental data. There are several possible explanations for this discrepancy. In the calculation, the radial orbital wave functions were optimised for the  W$^{14+}$ ground configuration. If the optimisation is performed for the ground configuration of the ionised W$^{15+}$ ion, the total cross section turns out to be in better agreement with the experimental data. The difference between the two different theoretical cross sections might be regarded as the inherent uncertainty of the present theoretical approach. Another source of uncertainty is the initial distribution among the various levels of the W$^{14+}$ ground configuration which are all sufficiently long lived such that they most probably all had been present in the ion beam of the experiment. Since the calculated EA cross sections are different for the individual initial levels (figure \ref{fig3}) the total cross section depends on the initial level distribution. Here we have assumed a statistical population. However, it might have been different in the experiment of Schury \etal \cite{Schury2019}.

\section{\label{conclusion}Conclusions}

In summary, we have performed a detailed comparison between the SCADW and LLDW methods for calculating the cross section for EISI of W$^{14+}$ ions. Significant differences between the two methods have been found for the $4d\to 5d$ EA channel, especially, near the ionisation threshold. In the LLDW calculation, the excited $4d^{9}4f^{12}5s^{2}5d$ configuration splits into a large number of levels that span a wide energy range straddling the ionisation threshold. The SCADW method treats this situation only  poorly and, therefore, severely underestimates the cross section of this EA channel. For the EA channels where the excited configurations do not straddle the ionisation threshold, there are only minor differences between the SCADW and LLDW cross sections. Consequently, our final total cross section for EISI of W$^{14+}$ ions is a hybrid of both approaches where the $4d\rightarrow 5d$ EA cross section is calculated with the LLDW approach while all other EA channels are evaluated using the less costly SCADW method. This hybrid cross section agrees much better  with the experimental results than the pure SCADW cross section. We expect that our approach will yield an improved  agreement between theoretical and experimental EISI cross sections also for the neighbouring charge states of tungsten ions for which significant discrepancies between measured and near-threshold SCADW  cross sections have been reported recently \cite{Schury2019}. As compared to full LLDW calculations the present hybrid approach is much less costly which is  particularly relevant for complex atomic systems that occur in technical applications such as magnetically confined  nuclear fusion.

\section*{Acknowledgments}

This work is supported by National Science Foundation of China (grant no. 11374365). Financial support by the German Federal Ministry of Education and Research (BMBF) within the \lq\lq{}Verbundforschung\rq\rq\ funding scheme (grant no.\ 05P19RGFA1) is gratefully acknowledged. B.E. is financially supported by a grant provided within the frame of the formal cooperation between the GSI Helmholtzzentrum f\"ur Schwerionenforschung (Darmstadt, Germany) and the Justus-Liebig-Universit\"at Gie{\ss}en.

\section*{References}


\providecommand{\newblock}{}

\end{document}